\documentclass[camera,letterpaper,nomarginnotes,nonarrowgutter]{jpaper}

\usepackage{listings}
\usepackage{multirow}
\usepackage{booktabs}
\usepackage[sort,compress]{cite}
\usepackage{amsmath}
\usepackage{graphicx}
\usepackage[bookmarks=true,breaklinks=true,letterpaper=true,colorlinks,linkcolor=black,citecolor=black,urlcolor=blue]{hyperref}
\usepackage{fancyhdr}
\usepackage{xcolor}

\newenvironment{acks}{\par\addvspace{\bigskipamount}%
  \noindent{\large\bfseries Acknowledgments}\par\nobreak\medskip\noindent\ignorespaces}{\par}

\definecolor{lstkeyword}{rgb}{0.10,0.10,0.70}
\definecolor{lstcomment}{rgb}{0.30,0.55,0.30}
\definecolor{lststring}{rgb}{0.65,0.15,0.15}
\definecolor{lstnumber}{rgb}{0.45,0.45,0.45}
\fancypagestyle{plain}{%
  \fancyhf{}%
  \fancyfoot[C]{\thepage}%
}

\begin{document}

\title{Ramulator 2.1: A Composable Memory System Simulator\\for Modern DRAM Systems}
\author{
{Haocong Luo}\qquad%
{F. Nisa Bostanc\i{}}\qquad%
{Ataberk Olgun}\qquad%
{Maria Makeenkova}\\%
{Ziad Malik}\qquad%
{Ipek Akdeniz}\qquad%
{Onur Mutlu}\\[2mm]
{\emph{SAFARI Research Group}}\\%
{\emph{ETH Z{\"u}rich}}}
\maketitle
\begin{abstract}
Ramulator 2.1 is a major overhaul of Ramulator 2.0 that substantially improves the simulator in three directions: 1)~support of modern and emerging DRAM and memory-controller features, 2)~better usability and extensibility of the simulator, and 3)~more comprehensive tests and validation workflows. Ramulator 2.1 adds support for advanced features in recent and emerging DRAM standards and memory controllers, including HBM3/4, LPDDR5/6, and GDDR7. To improve usability and extensibility, Ramulator 2.1 introduces a Python-based modeling and configuration interface backed by a two-way code-generation framework that 1) hides low-level C++ code behind high-level DRAM specifications written in Python, and 2) automatically creates Python proxies for all components of the simulator. Doing so enables users to rapidly create variants of DRAM standards and automate design-space-exploration workflows. To improve trustworthiness in simulation results, Ramulator 2.1 provides a comprehensive testing and validation infrastructure that covers both 1) fine-grained validation of specific DRAM timing constraints and memory-controller scheduling behavior, and 2) system-level performance evaluation using latency--throughput curves. {To aid performance analysis and debugging, Ramulator 2.1 also includes an easy-to-use and high-performance DRAM command trace visualizer.} Ramulator 2.1 is open-source on GitHub and under active development.
\end{abstract}

\thispagestyle{plain}
\pagestyle{plain}

\section{Introduction}
Ramulator 2.1~\cite{ramulator2github} is a major overhaul of Ramulator 2.0~\cite{kim2016ramulator, ramulatorgithub, luo2023ramulator2, ramulator2github, bostanci2026cleaning} that substantially improves 1) the coverage of modern DRAM~\cite{dennard1968dram} standards and memory-controller mechanisms, 2) the usability and extensibility of the simulator, and 3) the confidence users can place in simulation results. 

First, Ramulator 2.1 expands support for protocol-level features in recent and emerging DRAM standards and memory controllers that are absent in DDR4: \emph{parallel row/column command issue} in HBM3/HBM4~\cite{jedec_hbm3,jedec_hbm4} and GDDR7~\cite{jedec_gddr7}, \emph{split two-phase (ACT-1/ACT-2) activation} in LPDDR5/LPDDR6~\cite{jedec_lpddr5,jedec_lpddr6}, and \emph{data-clock synchronization} (WCK in LPDDR5/LPDDR6, RCK in GDDR7)~\cite{jedec_lpddr5,jedec_lpddr6,jedec_gddr7}. Ramulator 2.1 captures these features within a modular software framework that factors common mechanisms into a shared controller hierarchy and composable filtering predicates, letting each standard inject its protocol-specific scheduling constraints into a common command-selection pipeline while preserving standard-specific behavior. Such a design also enables future extensions of different DRAM request scheduling policies~{(e.g., \cite{kim2010tcm, mutlu2008PARBS, kim2010atlas, mutlu2007stall})}.

Second, Ramulator 2.1 introduces a Python-based {DRAM modeling and simulation configuration} interface supported by a two-way code-generation framework. This framework automatically 1) translates high-level DRAM specifications authored in Python into low-level C++ implementations compatible with Ramulator's cycle-level simulation methodology~\cite{kim2016ramulator, ramulatorgithub, luo2023ramulator2, ramulator2github} and 2) generates Python proxies {and exposes configurable parameters} for all other C++ simulator components. This design enables users to leverage Python's expressiveness and scriptability to 1) easily create variants of existing DRAM standards (e.g., adding a new DRAM command), and 2) streamline and automate experiment workflows for design space exploration and evaluation.

Third, Ramulator 2.1 strengthens the trustworthiness of the simulation results with a comprehensive testing and validation infrastructure. It supports both 1) fine-grained and low-level validation of specific DRAM timing constraints and memory-controller scheduling behavior, and 2) system-level performance evaluation using latency--throughput curves. {To aid performance analysis and debugging, Ramulator 2.1 also includes an easy-to-use and high-performance web-based DRAM command trace visualizer.}

Ramulator 2.1 is actively maintained, and we will continue incorporating new DRAM features and functionalities, including {Processing-Using-DRAM (PUD)~\cite{seshadri2015fast, seshadri2017ambit, hajinazar2021simdram, yuksel2024simultaneous, deoliveira2024mimdram, yuksel2024functionally, mutlu2024memory, olgun2022pidram, Oliveira2025Proteus, kim2018dram, kim2019drange, olgun2021quactrng, gao2019computedram, seshadri2013rowclone, baser2026simrapuf, yuksel2025trng, tokuda2026clutch, tokuda2026pudghost, kubo2024bulk, seshadri2020indrambulkbitwiseexecution, deng2018dracc, xin2020elp2im, shin2024prada, wong2026darthpum, truong2021racer, truong2026memory, gao2022fracdram, yuksel2025pudhammer} support} and {DRAM-based Processing-In-Memory (PIM)~\cite{lee2021hardware, he2020newton, kwon2022system, lee2022accelerator, farmahinifarahani2015nda, ahn2015scalable, ahn2015pim, nai2017graphpim, asghari2016chameleon, azarkhish2016logic, boroumand2018google, shin2018mcdram, zhang2014toppim, boroumand2021google, he2025papi, gu2025pim, park2024attacc, cho2020mcdramv2, oliveira2021damov, gomezluna2022benchmarking, devaux2019upmem}}, as the project evolves. We very much welcome feedback and contributions from the broader community, including both academia and industry. To this end, please use our GitHub repository~\cite{ramulator2github}, where we actively monitor and handle any raised issues.

\section{Modern DRAM and Controller Features}
\label{sec:features}

Modern DRAM standards introduce {key new features at the controller-device protocol level to enable high-performance and power-efficient operation. For example: 1) HBM3/4~\cite{jedec_hbm3,jedec_hbm4} and GDDR7~\cite{jedec_gddr7} allow \emph{parallel row/column command issue}, 2) LPDDR5/6~\cite{jedec_lpddr5,jedec_lpddr6} have \emph{split two-phase activation}, and 3) LPDDR5/6~\cite{jedec_lpddr5,jedec_lpddr6} and GDDR7~\cite{jedec_gddr7} introduce \emph{data-clock synchronization} (WCK and RCK).} 

{To implement these features in a clean and modular way, Ramulator 2.1 first implements a common abstract DRAM controller base class with {a} high-level memory request scheduling workflow. Second, Ramulator 2.1 extends this base class into different controller classes that implement different features with only {small additions and/or modifications} on top of the base workflow. To facilitate the clean and modular implementation, Ramulator 2.1 allows passing \emph{filtering predicates} in the form of lambdas into the base workflow to let each controller class inject feature-specific scheduling constraints.}

\noindent\textbf{Parallel row/column command issue.}
HBM3, HBM4, and GDDR7 provide separate {C/A (i.e., command/address) buses} for row commands
(e.g., \texttt{ACT}) and column commands
(e.g., \texttt{RD}, \texttt{WR}), allowing simultaneous issue of row and column commands. Ramulator 2.1 implements this feature by calling the base scheduling workflow \emph{twice}, each with a different filtering predicate that selects only row or column commands.

\noindent\textbf{Split ACT-1/ACT-2 activation.}
LPDDR5 and LPDDR6 split activation into two commands: 1) \texttt{ACT-1} that transitions a bank to an intermediate \emph{Activating}
state, and 2) \texttt{ACT-2} which must follow \texttt{ACT-1} within the deadline of $t_\text{AAD}$ cycles to
complete the activation. {To enable correct per-bank \texttt{ACT-2} issuing and deadline tracking, Ramulator 2.1's LPDDR5/6 controller passes filtering predicates that 1) only {allow} requests that have already issued \texttt{ACT-1} to issue \texttt{ACT-2}, and 2) {prevent} other requests from interrupting the pending issue of \texttt{ACT-2}.}

\noindent\textbf{WCK/RCK data-clock synchronization.}
To save power, the faster data transfer clocks of LPDDR5/6~\cite{jedec_lpddr5,jedec_lpddr6} (WCK) and GDDR7~\cite{jedec_gddr7} (RCK) can be enabled only when there is data transfer. {To do so, the controller must issue the corresponding synchronization commands (i.e., \texttt{CAS\_RD}/\texttt{CAS\_WR} for LPDDR5/6 and \texttt{RCKSTRT}/\texttt{RCKSTOP} for GDDR7) before the next read/write accesses. Ramulator 2.1 models this feature by explicitly tracking the data-clock active and expiry {windows} and injecting the required synchronization commands as prerequisites when needed, while preserving normal timing and scheduling constraints.}

\noindent\textbf{Other controller features.}
{Filtering predicates also facilitate the implementation of other controller features that change DRAM request and scheduling behavior. For example, Ramulator 2.1 implements BlockHammer~\cite{yaglikci2021blockhammer} and {PRAC~\cite{jedec2024ddr5, canpolat2025chronus, canpolat2024understanding}} controllers that use filtering predicates to 1) defer unsafe activation commands, and 2) ensure that ordinary requests do not interfere with required recovery or mitigation commands. We plan to implement many more {DRAM request scheduling policies~{(e.g., \cite{kim2010tcm, mutlu2008PARBS, kim2010atlas, mutlu2007stall, subramanian2016bliss,subramanian2014blacklisting, usui2016dash, ausavarungnirun2012staged, kim2012salp, yaglikci2022hira, chang2014improving})} in the future.}}

\section{Python Interface}
\label{sec:python}

Ramulator 2.1 provides a {straightforward and easy-to-use} Python interface that serves two purposes:
\emph{configuring simulations} and \emph{authoring DRAM standards}. A code generation framework \emph{automatically} {1) generates Python proxies and exposes configurable parameters and knobs for the actual Ramulator 2.1 code implemented in C++ so that {a} simulation can be flexibly configured and driven in Python}, and 2) translates high-level and human readable DRAM specifications in Python into low-level C++ code for Ramulator's cycle-level and hierarchical DRAM modeling and simulation methodology~\cite{kim2016ramulator, ramulatorgithub, luo2023ramulator2, ramulator2github} that reduces repetitive and boilerplate code exposed to the user.
\subsection{Simulation Configuration}
\label{sec:python-config}
{Ramulator 2.1 comes with a Python package {that} consists of Python proxy classes for each of the simulator components (e.g., frontend, controller, scheduler, address mapper) in the C++ codebase. Each Python proxy class contains the same set of configurable parameters as the corresponding C++ components. {With the Python package, users can flexibly instantiate and compose these proxy classes to configure the simulated system. Users can then run the simulation and collect structured statistics, all from a single Python script.} The Ramulator 2.1 Python package can also be seamlessly integrated with other simulators with a Python-based configuration system like gem5~\cite{Binkert2011gem5, lowepower2020gem520}. The Python package and all proxy classes are \emph{automatically generated} from the C++ source code during the build process, without the need {for} any manual maintenance (i.e.,  {transparently} to the user).}

{Under the hood, the Python proxies are lightweight and structured objects that only {mirror} the hierarchy and store the configurations of each simulator component \emph{without} any binding to the actual C++ object.} Doing so allows other simulators that do not use Python to configure the simulation to also integrate Ramulator 2.1 \emph{without} needing to include extra Python dependencies. Ramulator 2.1 provides a tool to \emph{automatically} generate an \emph{equivalent} pure text YAML configuration file from the Python configuration script that can be directly parsed by the C++ codebase.

\subsection{Authoring and Extending DRAM Standards}
\label{sec:python-dram}
{The Python package of Ramulator 2.1 also provides a high-level, human-readable, and easy-to-extend interface for authoring DRAM standards. The DRAM specifications (e.g., the DRAM organization hierarchy, supported DRAM commands, timing constraints) are described as plain Python dictionaries and lists of strings and numbers that directly {mirror} the names and values of DRAM organization, commands, and timing constraints, etc. The code generation framework then \emph{automatically} generates the low-level C++ code following Ramulator's cycle-level and hierarchical DRAM modeling and simulation methodology~\cite{kim2016ramulator, ramulatorgithub, luo2023ramulator2, ramulator2github}, hiding complicated, repetitive and boilerplate code from the user. For example, in Ramulator 2.1, the lines of code needed for DDR5~\cite{jedec2024ddr5} reduce by 67.2\% compared to Ramulator 2.0~\cite{luo2023ramulator2, ramulator2github} (from 402 lines of C++ code down to 132 lines of Python code).}

The Python-based DRAM specification authoring interface significantly simplifies the process of creating variations of DRAM standards by adding new DRAM commands and related timing constraints. Leveraging the flexibility and expressiveness of Python, to add a new DRAM command and timing constraints to a DRAM standard, users only need to 1) inherit from the base DRAM standard, and 2) append only the new commands and timings to the list of supported commands and timing constraints in the specification. For example, Listing~\ref{lst:ddr5_vrr} shows how to add a {Victim Row Refresh~\cite{kim2014flipping}} (VRR) command and its associated timing constraints on top of DDR5 with only 18 lines of Python source code (excluding blank and comment lines; compared to fully duplicating the entire 445 lines of C++ code in Ramulator 2.0). Table~\ref{tab:ramulator-dram-sloc} reports the reduction in lines of source code  (excluding blank and comment lines) across the DRAM standards and command variants currently implemented in Ramulator 2.1.

\begin{figure}[t]
\begin{lstlisting}[
  language=Python,
  numbers=left,
  caption={Extending DDR5 with a new {Victim Row Refresh} (VRR) command and its associated timing constraints.},
  label={lst:ddr5_vrr},
]
import math

from ramulator.dram.ddr5 import DDR5
from ramulator.dram.spec import TimingConstraint

# Inherit from DDR5
class DDR5_VRR(DDR5):
    name = "DDR5_VRR"
    # Append the new VRR command
    commands = DDR5.commands + ["VRR"]
    # Append the new timing constraints related to VRR
    timing_params = DDR5.timing_params + ["nVRR"]
    timing_constraints = DDR5.timing_constraints + [
        TimingConstraint(level="Bank", preceding=["VRR"], following=["ACT"], latency="nVRR"),
        TimingConstraint(level="Bank", preceding=["ACT"], following=["VRR"], latency="nRC"),
        TimingConstraint(level="Rank", preceding=["PREpb", "PREab"], following=["VRR"], latency="nRP"),
    ]


# Reuse all DDR5 presets
DDR5_VRR.org_presets = DDR5.org_presets
DDR5_VRR.timing_presets = {}
# Add the new nVRR timing constraint to all DDR5 presets
for _name, _timings in DDR5.timing_presets.items():
    _vrr_timings = dict(_timings)
    _vrr_timings["nVRR"] = math.ceil(280_000 / _timings["tCK_ps"])
    DDR5_VRR.timing_presets[_name] = _vrr_timings
\end{lstlisting}
\end{figure}

\begin{table}[t]
  \centering
  \small
  \renewcommand{\arraystretch}{1.05}
  \caption{Reduction in lines of source code of DRAM standards in Ramulator~2.1 (Python) compared to Ramulator~2.0 (C++).}
  \label{tab:ramulator-dram-sloc}
  \begin{tabular*}{\columnwidth}{@{}l@{\extracolsep{\fill}}c@{\hspace{1.5em}}c@{\extracolsep{\fill}}c@{}}
    \toprule
    \multirow{2}{*}[-0.75ex]{\shortstack[l]{\textbf{DRAM}\\\textbf{Standard}}} & \multicolumn{2}{c}{\textbf{Lines of Source Code}} & \multirow{2}{*}[-0.75ex]{\textbf{Reduction}} \\
    \cmidrule(lr){2-3}
     & \textbf{v2.0 (C++)} & \textbf{v2.1 (Python)} & \\
    \midrule
    DDR3      & 325 & 129 & 60.3\% \\
    DDR4      & 354 & 161 & 54.5\% \\
    DDR5      & 402 & 132 & 67.2\% \\
    GDDR6     & 327 & 199 & 39.1\% \\
    HBM1      & 287 & 133 & 53.7\% \\
    HBM2      & 289 & 146 & 49.5\% \\
    LPDDR5    & 395 & 143 & 63.8\% \\
    \midrule
    DDR4-VRR  & 375 & 18  & 95.2\% \\
    DDR5-VRR  & 445 & 18  & 96.0\% \\
    \midrule
    \textbf{Total} & \textbf{3,199} & \textbf{1,079} & \textbf{66.3\%} \\
    \bottomrule
  \end{tabular*}
\end{table}
\section{Tests and Validation}
{To improve the trustworthiness of Ramulator 2.1's modeling and simulation results, we implement a comprehensive testing and validation infrastructure that covers both 1) fine-grained and low-level validation of specific DRAM timing constraints and memory-controller scheduling behavior, and 2) system-level performance evaluation using latency--throughput curves.}

\noindent\textbf{Fine-Grained Validation.} We implement a unit test framework based on pytest~\cite{pytest} that allows users to easily 1) create a device (DRAM or controller) under test, 2) send DRAM commands or memory requests to the device, and 3) probe and check the internal states (e.g., prerequisite commands, timings) at specific cycles. Listing~\ref{lst:device_test} shows an example unit test that verifies a read command to a DDR4 device is blocked until the row is activated and the \texttt{nRCD} timing constraint is satisfied. We aim to keep improving the test coverage and ease of use of this framework as we keep developing and maintaining Ramulator 2.1, and we welcome contributions from the community.

\begin{figure}[t]
\begin{lstlisting}[
  language=Python,
  numbers=left,
  caption={An example fine-grained unit test for DDR4.},
  label={lst:device_test},
]
import pytest

import ramulator
import tests.device_timings.harness as device_timings

pytestmark = pytest.mark.device_timings

def dram_example_test():
    # Create the DRAM under test
    dram = ramulator.dram.DDR4(
      org_preset="DDR4_8Gb_x8", timing_preset="DDR4_2400R", rank=1
    )
    dut = device_timings.DeviceUnderTest(dram)

    addr = dut.addr_vec(Rank=0, BankGroup=0, Bank=0, Row=12, Column=0)

    # Probe the states of the DRAM for a RD command at cycle 0
    closed = dut.probe("RD", addr, clk=0)
    # Check: The prerequisite command is ACT.
    assert closed.preq == "ACT"
    # Check: Timing is OK here since no ACT has been issued yet!
    assert closed.timing_OK is True
    # Check: Not ready since the prerequisite is not met.
    assert closed.ready is False

    # Issue the ACT command at cycle 0.
    dut.issue("ACT", addr, clk=0)

    # Probe and Check: Before nRCD, the row state is correct for RD 
    # but timing still blocks it.
    early = dut.probe("RD", addr, clk=dut.timings["nRCD"] - 1)
    assert early.preq == "RD"
    assert early.timing_OK is False
    assert early.ready is False
    assert early.row_hit is True
    assert early.row_open is True

    # At nRCD, the same command becomes legal.
    ontime = dut.probe("RD", addr, clk=dut.timings["nRCD"])
    assert ontime.preq == "RD"
    assert ontime.timing_OK is True
    assert ontime.ready is True
\end{lstlisting}
\end{figure}

\begin{figure*}[!t]
  \centering
  \includegraphics[width=\textwidth]{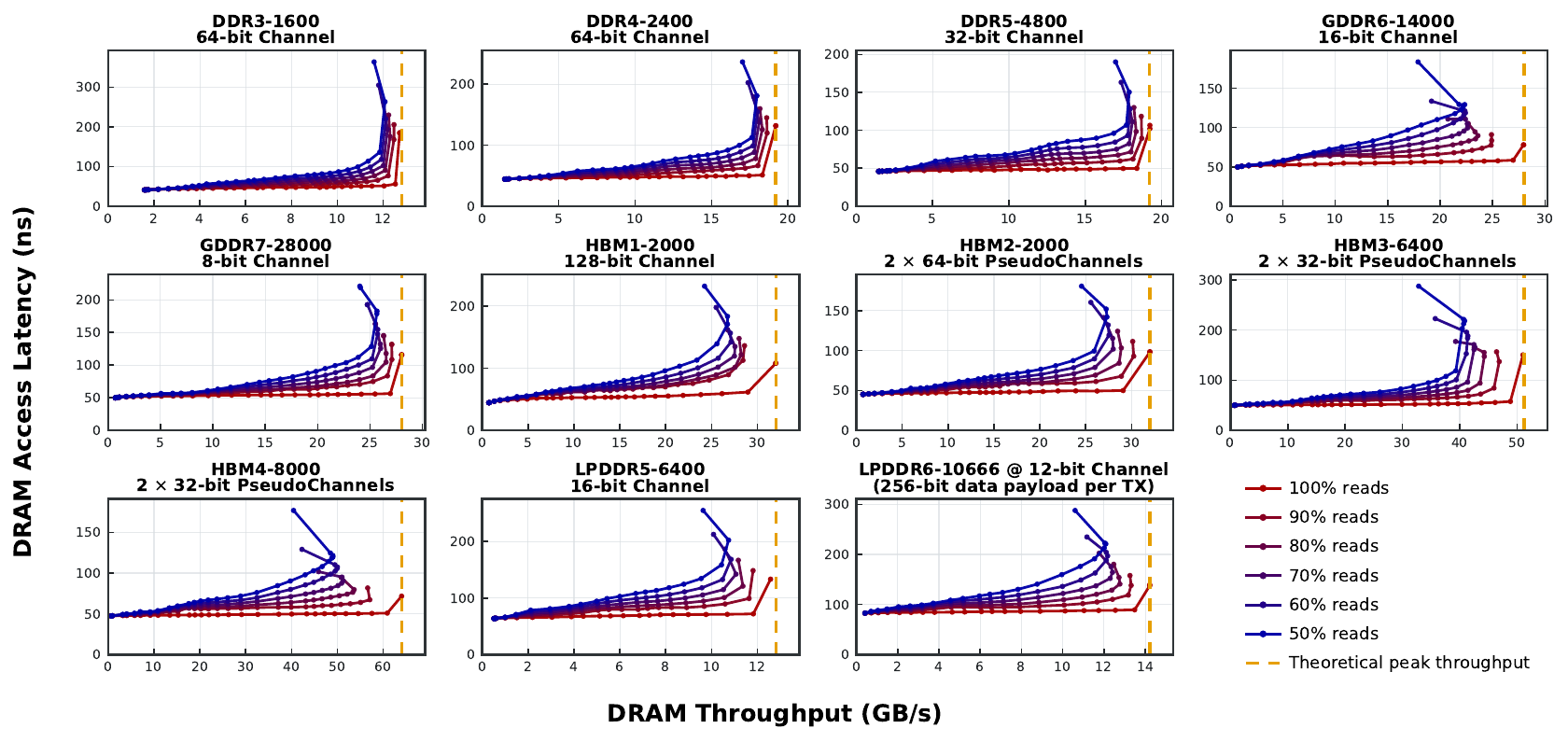}
  \caption[Latency--throughput curves across the DRAM standards modeled in
    Ramulator 2.1.]{Latency--throughput curves across the DRAM standards modeled in
    Ramulator 2.1.\protect\footnotemark}
  \label{fig:lat_tp}
\end{figure*}

\noindent\textbf{System-Level Performance Evaluation.} We perform high-level evaluation of the memory system performance by measuring the DRAM access latency under different loads (memory system throughput). To do so, we implement an improved version of the traffic-generator frontend from earlier work~\cite{bostanci2026cleaning} that generates two kinds of memory requests: 1) streaming requests with variable intervals between consecutive requests to adjust the load (throughput) of the memory system, and 2) serialized random-access requests that we use to probe the access latency (i.e., a probe request is issued only after the previous probe request completes). Figure~\ref{fig:lat_tp} shows the latency--throughput curves for all DRAM standards modeled in Ramulator 2.1. The y-axis shows the average access latency measured from the random-access probe requests, and the x-axis shows the throughput of the memory system. Different colors correspond to different read ratios (i.e., the fraction of read requests generated by the frontend, from 100\% reads, red, down to 50\% reads and 50\% writes, blue). We mark the theoretical peak throughput of each DRAM standard as vertical dashed orange lines. We observe that 1) for all DRAM standards, the simulated memory system can achieve the theoretical peak throughput, and 2) the latency--throughput curves follow the expected knee-curve shape. We conclude that Ramulator 2.1's modeling and simulation results are consistent with the expected behavior of real DRAM devices.
\footnotetext{This figure is updated to 1) fix a GDDR6 throughput issue, and 2) use better request depths for GDDR7 compared to the ICS'26 workshop version.}

\subsection{DRAM Command Trace Visualizer}
\begin{figure}[t]
  \centering
  \includegraphics[width=\columnwidth]{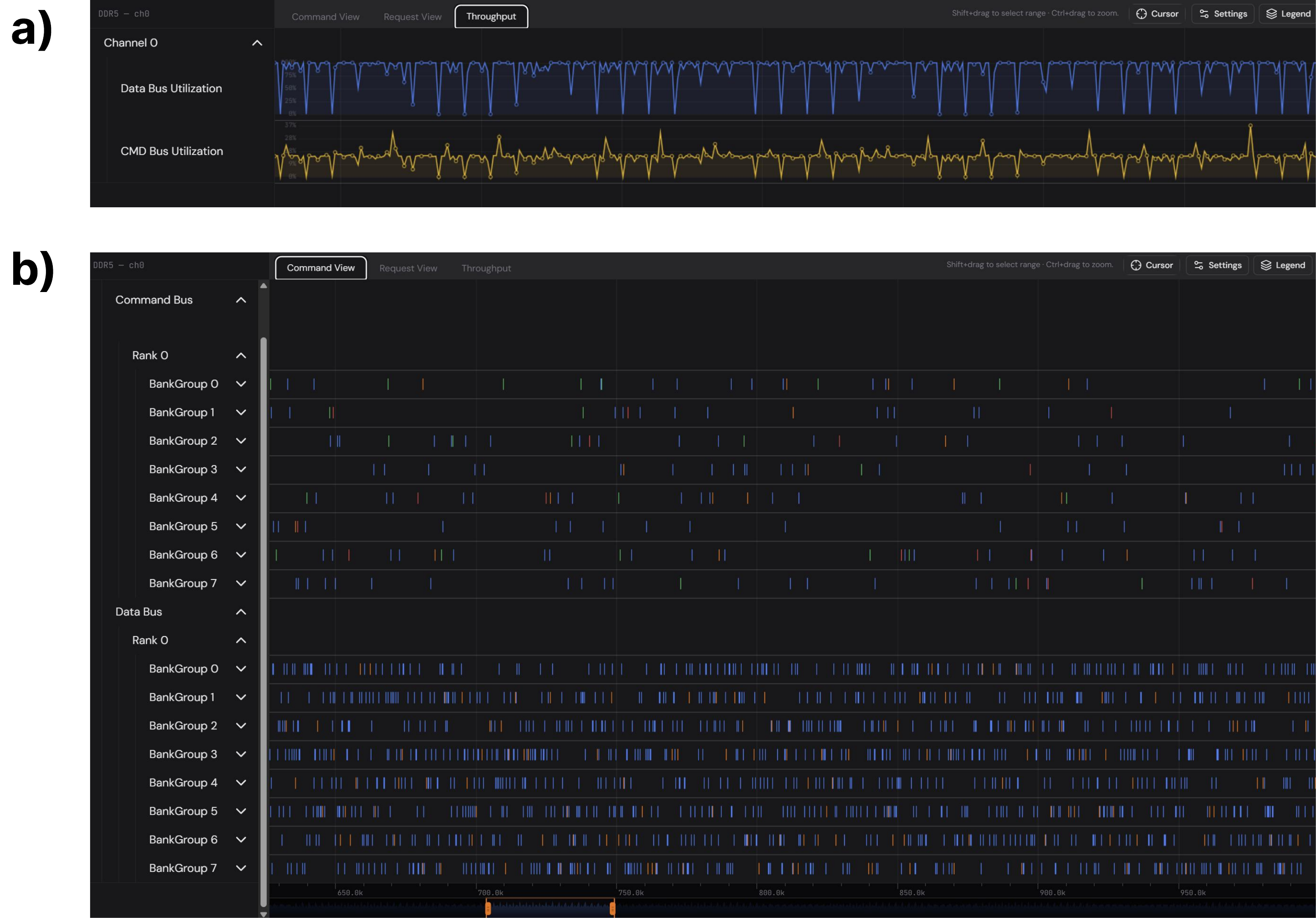}
  \caption{Ramulator 2.1's DRAM command trace visualizer: (a) the bus
    utilization view and (b) the command trace view.}
  \label{fig:visualizer}
\end{figure}

{To help users analyze DRAM system performance and debug the DRAM and memory controller models, Ramulator 2.1 includes a web-based, easy-to-use, and high-performance DRAM command trace visualizer. The visualizer can either 1) parse and visualize a recorded DRAM command trace offline, or 2) connect to a running Ramulator 2.1 simulation and visualize the DRAM commands being issued in real time. Users can easily inspect the DRAM command sequence, timings, addresses, and the address and data bus utilization interactively through the visualizer. Figure~\ref{fig:visualizer} shows two screenshots from the visualizer: (a) the bus utilization view illustrating the data bus and command bus utilization, and (b) the command trace view showing the sequence of DRAM commands and their addresses issued over time.}

\section{Conclusion}
We introduce Ramulator 2.1, a major overhaul of Ramulator 2.0 that improves the simulator along three
directions. First, it expands support for protocol-level features in recent and emerging DRAM
standards and memory controllers. Second, it introduces a Python-based DRAM modeling and simulation configuration interface, enabling users to rapidly create variants of DRAM standards and automate design-space-exploration workflows. Third, it strengthens the trustworthiness of simulation results with a comprehensive testing and validation infrastructure that spans fine-grained DRAM-timing and scheduling checks and system-level latency--throughput evaluation. {Our results show that Ramulator 2.1 significantly reduces the lines of code needed to implement DRAM standards compared to Ramulator 2.0. Ramulator 2.1 also provides latency--throughput curves that match curves expected from real systems for 11 DRAM standards}.

Ramulator 2.1 is open-source and under active development. We will continue to incorporate new DRAM features and functionalities. We are {very open} to and warmly welcome feedback and contributions from both academia and industry through our GitHub repository~\cite{ramulator2github} to improve Ramulator 2.1.

\begin{acks}
We thank the SAFARI Research Group members for their constructive feedback and for providing a stimulating intellectual{, scholarly,} and scientific environment. We acknowledge the generous gift funding provided by our industrial partners (especially Google, Huawei, Intel, Microsoft), which has been instrumental in enabling the research we have been conducting on read disturbance in DRAM in particular and memory systems in general~\cite{kim2014flipping, mutlu2017rowhammer,mutlu2019processing,mutlu2019rowhammer,mutlu2022modern,mutlu2023fundamentally,mutlu2023retrospective,mutlu2014research,mutlu2023retrospective-raidr,mutlu2023retrospectiveexperimentalstudydata, mutlu2025memory,mutlu2013memory, kakolyris2026columnkeeper, oliveira2021damov, gomezluna2022benchmarking, ghose2019processing, mutlu2024memory, mutlu2015main, oliveira2022accelerating, singh2021fpga, cai2017flashtbd, cai2017errors, mutlu2020intelligentdate, mutlu2023tesseractretrospective, mutlu2023selfretrospective, yuksel2026memory, olgun2026drambender, bostanci2026extended}. This work was in part supported by a Google Security and Privacy Research Award and the Microsoft Swiss Joint Research Center.
\end{acks}

\bibliographystyle{IEEEtran}
\bibliography{refs}

\appendix
\section{Errata}
\label{app:errata}

Figure~\ref{fig:lat_tp} is updated from the ICS'26 workshop version to fix a
GDDR6 throughput issue and to use better request depths in the memory controller for GDDR7.

\end{document}